\documentclass{iau} 
\usepackage{graphicx}

\title[SNR-CR connection] %% give here short title %%
{Supernova Remnant-Cosmic Ray connection: a modern view}

\author[G. Morlino]   %% give here short author list %%
{Giovanni Morlino$^1$}

\affiliation{$^1$Gran Sasso Science Institute, viale F. Crispi 7, 67100 L'Aquila, Italy. 
\\email: {\tt giovanni.morlino@infn.gssi.it}}

\pubyear{2015}
\volume{331}  %% insert here IAU Symposium No.
\jname{SN 1987A, 30 years later}
\editors{A.C. Editor, B.D. Editor \& C.E. Editor, eds.}
\begin{document}

\maketitle

\begin{abstract}
The Cosmic Ray (CR) physics has entered a new era driven by high precision measurements coming from direct detection (especially AMS-02 and PAMELA) and also from gamma-ray observations (Fermi-LAT). 
In this review we focus our attention on how such data impact the understanding of the supernova remnant paradigm for the origin of CRs.
In particular we discuss advancement in the field concerning the three main stages of the CR life: the acceleration process, the escape from the sources and the propagation throughout the Galaxy. We show how the new data reveal a phenomenology richest than previously thought that could even challenge the current understanding of CR origin.
%% add here a maximum of 10 keywords, to be taken form the file <Keywords.txt>
\end{abstract}

\firstsection % if your document starts with a section,
                   % remove some space above using this command.

%SECTION%
\section{Introduction}		\label{sec:intro}

Understanding the origin of Galactic cosmic rays (CR) is a puzzle consisting of three main pieces that must fit in with each other to produce a coherent picture and can be summarized as follow: 1) finding the main sources able to accelerate particles up to the highest energy we observe; 2) understanding how particles escape from their sources and are released into the interstellar medium (ISM);  3) understanding how particles propagate through the Galaxy before reaching the Earth. 
In this concise review I will outline some of the many recent observational findings that, to some extend, are challenging the current understanding of the Galactic CR origin. I will list those discoveries in connection to the three pillars outlined above.

The main sources of Galactic CRs are thought to be the shocks produced by the explosion of SNe, mainly because of energetic reasons, but also because the presence of accelerated particles is unequivocally inferred from the non-thermal radiation emitted by supernova remnants (SNR). The shock acceleration process is maybe the most studied mechanism related to the CR physics \cite[(see, e.g.,  Drury, 1983, and Blasi, 2013, for a review)]{Drury:1983,Blasi:2013}. In \S~\ref{sec:acceleration} we will focus our attention mainly on the particle spectrum and the maximum energy predicted in the framework of Diffusive Shock Acceleration (DSA), showing how recent observations of SNRs in the $\gamma$-ray band are challenging the theoretical predictions.

The final spectrum released into the Galaxy by SNRs is determined by the escaping process and could be significantly different from the spectrum accelerated at the shock.
Compared to the acceleration phase, the process of particle escaping has received much little attention, partially because the theoretical difficulty in describing how particles can reach their maximum energy and how they propagate in the transition region between the shock and the undisturbed ISM, where the average level of Galactic CRs dominates over the flux coming from a single source. In addition there are observational difficulties in detecting signals from this transition region with a noticeable exception: the detection of $\gamma$-ray emission from molecular clouds (MC) locate close enough to SNRs. The interesting it such kind of systems has been renewed recently thanks to the measurements of the ionization degree in these same clouds which can provide information on the CR spectrum below the threshold of the $\gamma$-rays production, paving the way to the study of CR spectrum from MeV up to TeV energies, and possibly beyond. This argument will be discussed in \S~\ref{sec:propagation}.

Finally, the study of the propagation mechanism through the Galaxy is receiving a lot of attention in the very recent years, thanks to the impressive amount of data coming from direct measurements of the CR flux and to information coming from the diffuse Galactic $\gamma$-ray background. The interpretation of these new data probably requires a revision of the current understanding of the propagation process and could even shed light on the role that CRs have in the evolution of the Galaxy. We will discuss such topics in \S~\ref{sec:propagation}.

%SECTION%
\section{Acceleration phase}	\label{sec:acceleration}

\subsection{Gamma-ray emission from SNRs}    \label{sec:gamma}
It is beyond any doubt that SNRs can accelerate CRs. The question is rather which is the total amount of energy channeled into  relativistic particles and which is the final spectrum injected into the ISM. $\gamma$-ray observations provide a privileged tool to answer these questions, allowing to directly infer the properties of accelerated hadrons, something that cannot be done in any other wavelength.

It is well know that the $\gamma$-ray emission can be either produced by hadronic processes ($\pi^0$ decay from hadronic collisions) or by leptonic one (inverse Compton -- IC) and often it is not easy to distinguish between the two processes. On a general ground, the detection of hadronic emission would favor the high efficient acceleration scenario, while the leptonic interpretation forces one to assume a quite low acceleration efficiency. This binomial is mainly due to the different  magnetic field strength required: IC scenario usually needs very low magnetic field ($\sim 10 \mu$G, comparable with the ISM value)  in order to simultaneously account for radio, X-ray and $\gamma$-ray emission. On the contrary, hadronic scenarios requires much larger values, of the order of few hundreds $\mu$G, that cannot result from the simple compression of interstellar magnetic field, but requires some sort of amplification. Indeed, the magnetic field amplification is thought to be a manifestation itself of an efficient acceleration process (see \S~\ref{sec:Emax}).

I real cases it is not always easy to distinguish between leptonic and hadronic scenarios. The case of RX~J1713 is very significative to this respect. This remnant has been considered for long time the best candidate for an efficient acceleration scenario, mainly due to its high $\gamma$-ray luminosity. The detection of $\gamma$-ray emission in the range [1-300 GeV] by the Fermi-LAT satellite have shown an unexpected hard spectrum which, at a first glance, seems to be more in agreement with a leptonic scenario. 
Nevertheless, a deeper analysis shows that neither the hadronic nor the leptonic scenarios, taken in their simplest form, can unequivocally explain the observations \cite[(Morlino \etal, 2009; Gabici \& Aharonian, 2016)]{Morlino+2009, GabiciAharonian2016}, hence the issue remains open.

%------------------------------------------------------------------------
\begin{figure}[t]
% \vspace*{-2.0 cm}
\begin{center}
 \includegraphics[width=3.4in]{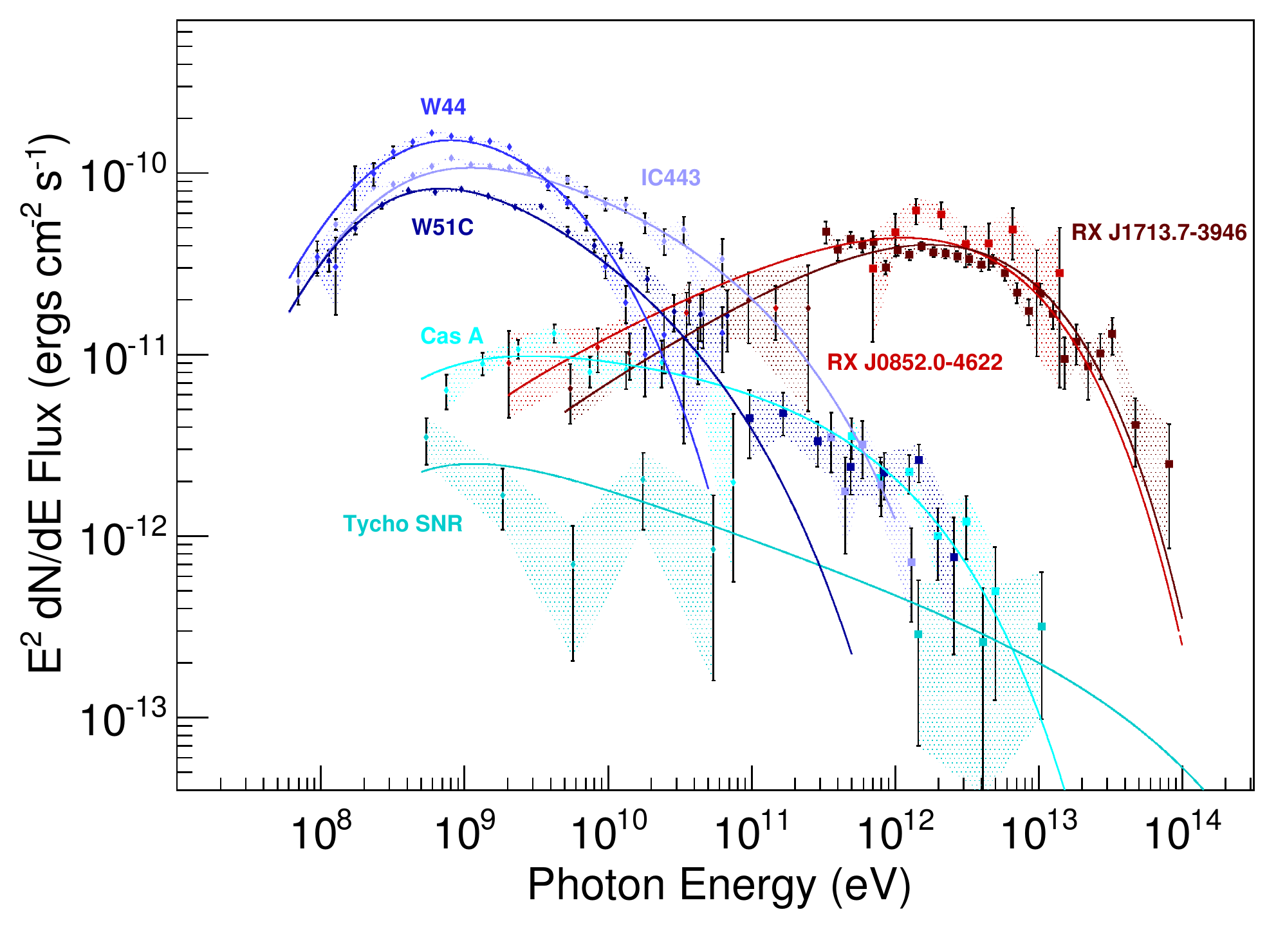} 
% \vspace*{-1.0 cm}
 \caption{$\gamma$-ray spectrum detected from some SNRs. The color code characterizes the age of the remnants: very young (Tycho and Cas A, $t\lesssim 400$ yr), young (RX J0852 and RX J1713, $t \simeq 1000\div2000$ yr) and middle age (W51C, W44 and IC443, $t\simeq 1\div2 \times 10^4$ yr). Figure taken from \cite{Funk2015}.}
   \label{fig:SNRspectra}
\end{center}
\end{figure}
%------------------------------------------------------------------------

From a theoretical point of view the diffusive shock acceleration (DSA) predicts a power low in momentum $\propto p^{-4}$ (which translates into an energy spectrum $\propto E^{-2}$ for relativistic energies). A remarkable property of DSA is that this prediction turns out to be very solid: even when the complex chain of non-linear effects is taken into account, the deviation from a straight power law are very mild. 
On the other hand, observations shows a much more varied phenomenology. The spectra inferred from $\gamma$-ray observations ranges from $E^{-2.8}$ up to $E^{-1.5}$, with an average slope $\approx -2.2\div -2.3$.
Some authors have suggested the presence of a general trend of the $\gamma$-ray spectrum with the age \cite[see Funk, 2015]{Funk2015}. The idea is summarized in the Figure~\ref{fig:SNRspectra}: very young SNRs (still in the free expansion phase) show a slight steep power law with index $\approx E^{-2.2} \div E^{-2.4}$ (like Tycho and Cas A); young SNRs ($t \approx 1000-2000$ yr) shows a quite hard spectrum more easily explained in terms of leptonic emission (but see the caveat above); middle aged SNRs show spectra of undoubtedly hadronic origin\footnote{Two middle aged SNRs, IC443 and W44, also show the characteristic pion bump around $\approx 280$ MeV, un unmistakable signature of hadronic origin of $\gamma$-rays \cite[(Ackermann \etal, 2013)]{Ackermann+2013}.}, steeper than $\propto E^{-2.5}$. Moreover the first two categories have been detected up to tens of TeV, implying a maximum energy of primary protons up to $\lesssim 100$ TeV (maybe one order of magnitude less for Tycho and Cas A), while the middle aged SNRs have been detected up to much lower energies.
Nevertheless, it is hard to say whether such a trend is intrinsic in the SNR evolution or if it is rather due to environmental effects. It has been shown, for example, that the presence of a non negligible fraction of neutral hydrogen can significantly modify the shock structure through the process of charge exchange and ionization, resulting in spectra steeper than $E^{-2}$ \cite[(Morlino \etal, 2010, Morlino \& Blasi, 2016)]{Morlino+2013, MorlinoBlasi2016}. 
Another complication may be produced by the presence of dense clumps in the CSM \cite[(Gabici \& Aharonian, 2014)]{GabiciAharonian2014}. In this case, when the shock crosses the clumps, hydrodynamical instabilities can amplify the magnetic field all around the clump, making difficult for particles at low energies to penetrate inside the clump compared to the most energetic ones. The final result would be a $\gamma$-ray spectrum harder then the parent proton spectrum.

Finally, steeper spectra may also result from a more subtle aspect of the non linear DSA theory. If the magnetic field is efficiently amplified, the speed of the magnetic turbulence could be enhanced up to a non negligible fraction of the shock speed, hence reducing the effective compression factor felt by accelerated particles \cite[(Morlino \& Caprioli, 2012)]{MorlinoCaprioli:2012}. Unfortunately such mechanism dependents on plasma conditions which are very difficult to test.

\subsection{Maximum energy and magnetic field amplification}    
\label{sec:Emax}
The maximum energy required to explain the CR spectrum observed at Earth is around few PeV for protons and then it scales like the nuclear charge, for heavier nuclei. 
In the framework of DSA applied to SNRs, there is a general consensus on the fact that the maximum energy is reached before the beginning of the Sedov-Taylor (ST) stage. The reason is that particles diffusing ahead of the shock will be cached by the shock only if the shock is moving faster than the particle. The average displacement of a particle diffusing ahead of the shock is $(D t)^{1/2}$ while the shock radius moves like $R_s(t) \propto t^{\alpha}$ hence particles will be recaptured only if $\alpha > 1/2$. But in the ST phase $\alpha=2/5$, hence particles can escape, while in the free expansion phase $\alpha (\lesssim 1) > 1/2$ (the exact value of $\alpha$ slightly depends on the velocity profile of the ejecta and on the density profile of the circumstellar medium). This reasoning became more complicated if we try to account for the time dependence of the magnetic field amplification but, because we do expect the amplification to increase with increasing shock speed, the general feeling is that the maximum energy should stop increasing even before the end of the free expansion phase.
Now, the Sedov-Taylor phase begins at $t_{\rm ST} \approx 50 \, (M_{\rm eje}/M_{\odot})^{5/6}\, (E_{\rm SN}/10^{51} {\rm erg})^{-1/2} \, (n_{\rm ISM}/ \rm cm^{-3})^{-1/3}$ yr, which, for typical values of the parameters,  ranges between 50 an 200 years, while the acceleration time is given by $t_{acc} \approx 8 D/u_{sh}^2$. In the framework of linear DSA, the diffusion coefficient is given by $D = r_L v /(3 \mathcal{F})$, where $r_L$ is the Larmor radius of particles and $v$ their speed, while $\mathcal{F}(k)$ is the logarithmic power spectrum of magnetic waves with wave-number $k$. 
Equating the acceleration time with $t_{\rm ST}$ we get the following estimate for the maximum energy:
\begin{equation} \label{eq:E_max}
  E_{\max} = 5 \times 10^{13} \, Z \, {\cal F}(k_{\min}) \,
  		   \left( \frac{B_0}{\mu \rm G} \right)
                    \left( \frac{M_{\rm ej}}{M_{\odot}} \right)^{-\frac{1}{6}} 
                    \left( \frac{E_{\rm SN}}{10^{51} \rm erg} \right)^{\frac{1}{2}} 
		   \left( \frac{n_{\rm ISM}}{\rm cm^{-3}} \right)^{-\frac{1}{3}} \; {\rm eV} \,,
\end{equation}
where $k_{\min} = 1/r_L(E_{\max})$ is the wave number resonant with particles at maximum energy.
We notice that more realistic estimates of the maximum energy (for example accounting for the fact that the shock speed is slightly decreasing also during the ejecta-dominated phase) usually return somewhat lower values. 

Eq.(\ref{eq:E_max}) depends only weakly on the environmental parameters, while depends strongly on the level of magnetic turbulence. As a consequence, the maximum energy of protons could reach few PeV only if ${\cal F}(k_{\min}) \gg 1$, namely the magnetic turbulence at the scale of $r_L(E_{\max})$ must be much larger than the pre-existing field, i.e. $\delta B \gg B_0$. Clearly if this condition were realized, the linear theory used to derive the diffusion coefficient would not hold anymore. 
Apart from that, the value of turbulence in the ISM at scales relevant here is $\delta B/B_0 \lesssim 10^{-4}$ (\cite[(Armstrong \etal, 1981)]{Armstrong+1981},) hence, in absence of any mechanism able to amplify the magnetic turbulence, SNR shocks could accelerate protons only up to the irrelevant energy of a few GeV. 

This puzzle has been partially solved by the idea that the same accelerated particles can amplify the magnetic field upstream through the {\it resonant streaming instability} while they try to diffuse far away from the shock \cite[(see Drury, 1983, or Blasi, 2013, for a review)]{Drury:1983, Blasi:2013}. Nevertheless the resonant instability can only produce $\delta B \lesssim B_0$, i.e. ${\cal F} \lesssim 1$, resulting in a maximum energy for protons of $10-100$ TeV. Hence more effort is needed to fill up the last decade of energy to reach the PeV. The solution to this conundrum probably resides in other types of instabilities that CRs can excite, the most promising one being the {\it non-resonant Bell instability} \cite[(Bell, 2004)]{Bell2004}. This instability results from the $\bf{j} \times \bf{B}_0$ force that the current due to escaping particles produces onto the plasma and grows very rapidly for high Mach number shocks. However, the scales that get excited are very small compared with the gyration radii of accelerated particles. Hence, it is not clear if the highest energy particles can be efficiently scattered.
Indeed, hybrid simulations seem to confirm that the non-resonant Bell instability grows much faster than the resonant one for Mach number $\gtrsim 30$ and produces ${\cal F}\gg1$ \cite[(Reville \& Bell,2012; Caprioli \& Spitkovsky, 2014)]{Reville-Bell:2012,Caprioli-Spitkovsky:2014}. The same simulations also show that the instability produces a complex filamentary structure which could be able to scatter particles efficiently.

The efficiency of Bell instability is determined by the strength of the return current that balance the current of escaping CRs which, in turn, depends on the density of the circumstellar medium and on the shock speed. Indeed, \cite{Bell+2013} showed that the conditions to reach PeV energies are rather special and can be obtained only during the first few weeks or decades following the explosion of stars occurring in dense circumstellar winds. On the contrary, type Ia SNe, occurring in a less dense ISM, should be able to produce only $\sim 100$ TeV protons. A similar conclusion has been reached by \cite{Cardillo+2015}. 
If this scenario would be confirmed, it would not be a surprise the fact that we have not detected yet a single SNR acting as a PeVatron. The chances for such a discovery will be surely enhanced with the forthcoming Cherenkov facilities (CTA).

\begin{figure}[t]
 \vspace*{-1.8 cm}
\begin{center}
 {\includegraphics[width=5cm]{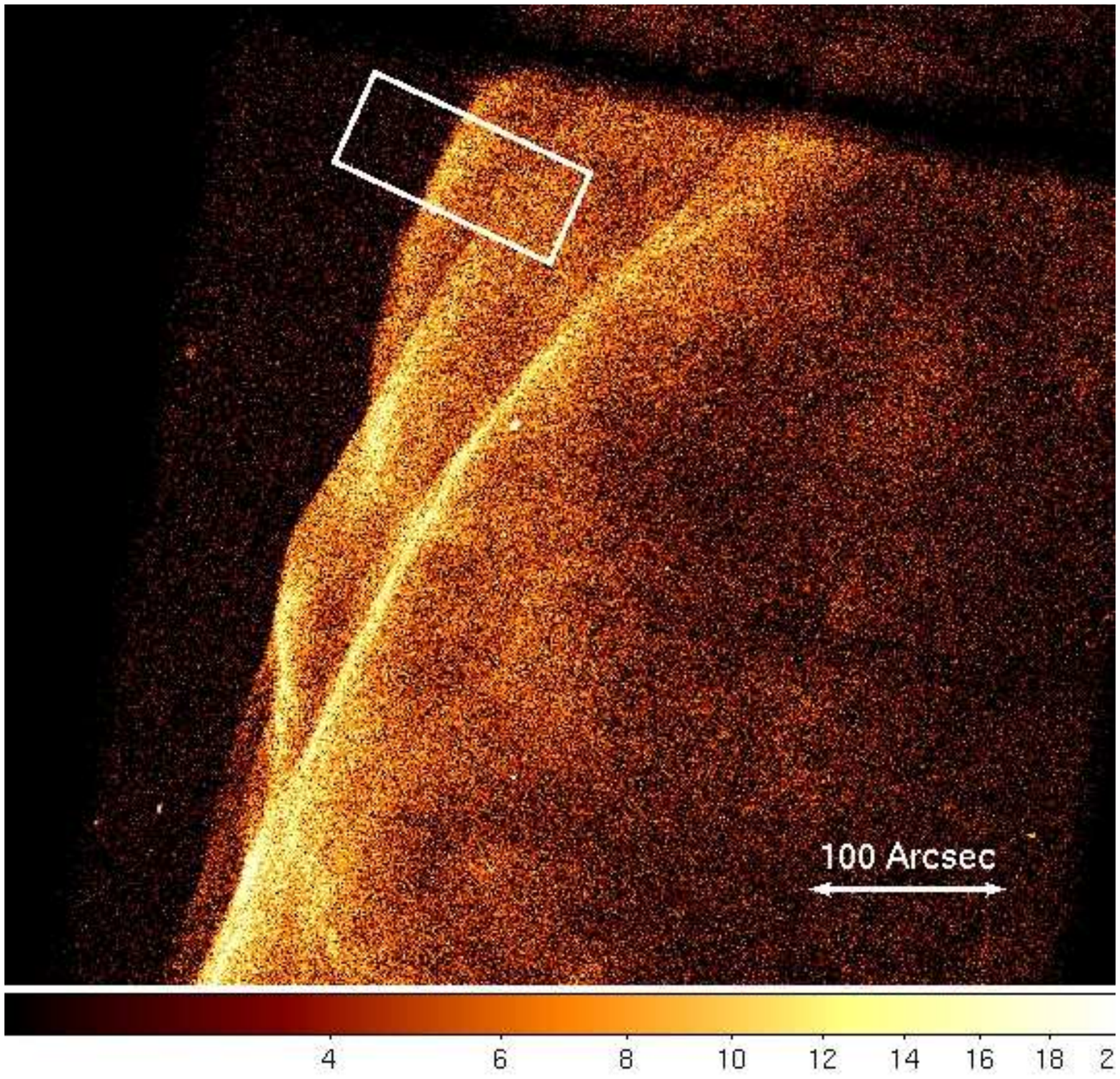} 
 \includegraphics[width=7cm]{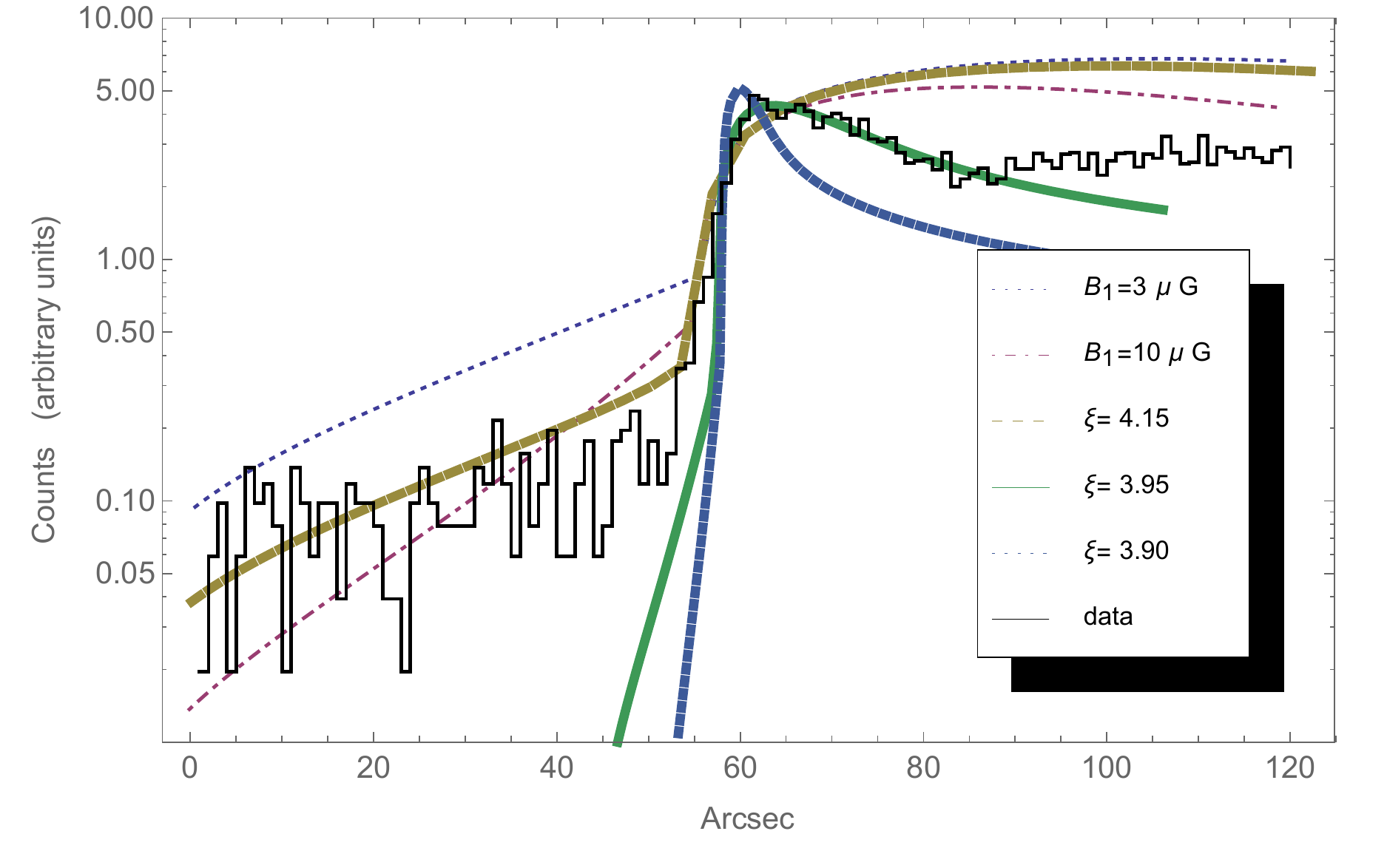}}
 \caption{Left: {\it Chandra} observation of the Northeastern shock of SN 1006. Right: Photon counts as a function of the position corresponding to the white box in the left picture. The curves represent theoretical predictions of non-thermal X-ray profile assuming different acceleration efficiency \cite[(Morlino \etal, 2010)]{Morlino+2010}. The plot shows that models without magnetic amplification also in the upstream region (thin curves)  over predict the X-ray emission from this region and, simultaneously, cannot account for the bump at the shock location.}
   \label{fig:SN1006}
\end{center}
\end{figure}

We conclude this section recalling that one of the most remarkable finding of the last decade is that large magnetic field strength, much larger than the average galactic one, have been inferred in almost all young SNRs, through the observation of thin X-ray filaments at the forward shocks \cite[(see Vink, 2012; Ballet, 2006, for recent reviews)]{Vink2012, Ballet2006}. This provide a strong support to the presence of some amplification mechanisms.
One should keep in mind, however, that the magnetic amplification can increase $E_{\max}$ only if it occurs both upstream and downstream of the shock otherwise particles could escape either from one side or the other. Having a magnetic amplification downstream only is quite an easy task: in fact, the shocked plasma is usually highly turbulent and hydrodynamical instabilities can trigger the amplification, converting a fraction of the turbulent motion into magnetic energy, as shown by \cite{Giacalone-Jokipii:2007}. Conversely, there are no reasons, in general, to assume that the plasma where a SNR expands is highly turbulent to start with.
Indeed, evidences that the amplification occurs also upstream have been collected from the  X-ray observations of SNR shocks \cite[(Morlino \etal, 2010; Eriksen \etal, 2011)]{Morlino+2010, Eriksen+2011} (see also Figure \ref{fig:SN1006}) confirming that CRs are the main agent to drive such a process.

%------------------------------------------------------------------------
%------------------------------------------------------------------------

%SECTION%
\section{Escaping from the sources}	\label{sec:escape}

In the test-particle picture of DSA, accelerated particles are advected downstream of the shock and will be confined in the interior of the SNR until the shock disappears and the SNR merges into the ISM. At that point particles will be released in the ISM but they would have lost part their energy because of the adiabatic expansion of the remnant: hence the requirements in terms of maximum energy at the source would be even more severe than they already are. 
Therefore, effective escape from upstream, while the acceleration is still ongoing, is fundamental if high energy particles must be released in the ISM. 
The description of how particle escape from a SNR shock has not been completely understood yet, the reason being the uncertainties related to how particles reach the maximum energies (a careful description of the numerous problems involved can be found in \cite[Drury, 2011]{Drury:2011}). Below we just describe the general framework.

Let us assume that the maximum momentum, $p_{\max,0}$, is, indeed, reached at the beginning of the ST phase as discussed in \ref{sec:Emax} and, afterwards, it drops with time as $p_{\max}(t) \propto (t/t_{\rm ST})^{-\beta}$, with $\beta > 0$. The energy in the escaping particles of momentum $p$ is
\begin{equation} \label{eq:f_esc}
  4 \pi f_{\rm esc}(p) \,pc \, p^2 dp = \xi_{\rm esc}(t) \, \frac{1}{2} \rho u_{\rm sh}^3 \, 4 \pi R_{\rm sh}^2 \, dt
\end{equation}
where $\xi_{\rm esc}(t)$ is the fraction of the incoming energy flux, $\frac{1}{2} \rho u_{\rm sh}^3 4 \pi R_{\rm sh}^2$, that is converted into escaping flux.
If the expansion occurs in a homogeneous medium with $R_{\rm sh} \propto t^{\alpha}$ and $V_{\rm sh} \propto t^{\alpha-1}$, therefore, since $dt/dp \propto t/p$, from Eq.~(\ref{eq:f_esc}) we have:
\begin{equation} \label{eq:N_esc}
  f_{\rm esc}(p) \propto p^{-4} t^{5\alpha-2} \xi_{\rm esc}(t).
\end{equation}
It follows that in the ST phase, where $\alpha=2/5$, the spectrum released in the ISM is $f_{\rm esc}(p) \propto p^{-4}$ only if $\xi_{\rm esc}$ keeps constant with time. It is worth stressing that this $p^{-4}$ has nothing to do with the standard result of the DSA in the test-particle regime. Neither does it depend on the detailed evolution in time of the maximum momentum. It solely depends on having assumed that particles escape the SNR during the adiabatic phase.
Notice also that in more realistic calculations $\xi_{\rm esc}$ should decreases with time, leading to a spectrum of escaping particles which is even harder than $p^{-4}$ \cite[(Caprioli \etal, 2010)]{Caprioli-AB:2010}. On the other hand, the total spectrum of particles injected into the ISM by an individual SNR is the sum of two populations: particles escaping during the acceleration and particles released after the shock dissipates. The latter component is not modified by the escaping process and should reflect closely the accelerated spectrum. 
An interesting byproduct of this scenario is that the intersection of the two populations may tentatively be identified with the change in spectral index at $\sim 300$ GeV detected by PAMELA and AMS-02  \cite[(Bell, 2015)]{Bell:2015}. We will discuss this feature more in details in the next section.

A further complication arises when one try to account for the environment when SNR expands. While type Ia SN are thought to expand in a pretty homogeneous ISM, core-collapse SNe expand into the wind/bubble system generated by the progenitor stars. In such a situation the remnant evolution differs significantly from the classical ST solution and could results in CR spectra different from the the standard $p^{-4}$.
In spite of such complications, it seems that when one sum the contribution of CRs from different type of SNRs, the final results approach $p^{-4}$ \cite[(Caprioli \etal, 2010)]{Caprioli-AB:2010}. Nevertheless one should take this result with caution, in that it is obtained extrapolating the effect of streaming instability calculated in the linear regime, to the non linear one, and neglecting the more complicated effect of the non-resonant instability, which should play a major role when the acceleration efficiency is high.
In conclusion, the theory does not provide yet a solid answer to the question of escaping from sources. Therefore, it is of the outermost importance to look for possible observational constraints that can help to understand such a process.

\underline{MC-SNR associations}. 
A possible way to investigate the escaping process is to look for $\gamma$-ray emission from MC-SNR associations. When MCs are located close enough to a SNR, escaping CRs can ``illuminate'' the cloud, producing $\gamma$-rays through interaction with the cloud gas. The resulting $\gamma$-rays can provide information on the escaping CR spectrum but also on the diffusion coefficient in the region between the SNR and the MC, which determines how fast particles escape \cite[(Gabici \etal, 2009)]{Gabici+2009}. 

MC-SNR systems could also be used to shed light on the lowest energy part of the CR spectrum, namely below the pion threshold of $\sim 280$ MeV, allowing us to test models on CR acceleration and propagation over an energy interval  spanning from the MeV to the TeV domain. Such impressive result can be reached looking for the ionization level of MC \cite[(see Gabici \& Montmerle, 2015, for a review)]{GabiciMontmerle2015}. 
In fact, even though CRs with energy below $\sim 280$ MeV carry a subdominant fraction of the bulk energy of CRs, they nevertheless play a pivotal role in regulating the properties of the interstellar medium, by ionising and heating the gas, and thus driving interstellar chemistry in MCs, where ionizing photons cannot penetrate \cite[(Dalgarno, 2006)]{Dalgarno2006}.  Moreover, being the main regulators of the ionization fraction inside MCs, low energy CRs set the level of coupling between magnetic field and gas and thus influence the process of formation of stars and planets.  Hence, finding observational ways to trace them and quantify these effects is of prime interest.
A remarkable result has been already achieved studying, among the others, the MC complex around the SNR W28, where the MC closest to the SNR shows a strong enhancement of the ionization rate \cite[(Vaupr\'e \etal, 2014)]{Vaupre+2014}  signature of the fact that low energy CRs has reached only the closest MC, while high energy CRs also reached the more farther MCs, producing the characteristic $\gamma$-ray emission.

%SECTION%
\section{Propagation through the Galaxy} \label{sec:propagation}

%------------------------------------------------------------------------
%\begin{figure}
%\begin{center}
% \includegraphics[width=6cm]{fig3_halo.eps} 
% \caption{Schematic representation of the Galactic diffusion model: CRs are produced by sources in the galactic disc and diffuse in the magnetic halo above and below the disc, before escaping in the intergalactic medium.}
%\label{fig:halo}
%\end{center}
%\end{figure}
%------------------------------------------------------------------------

After leaving the near-source region, CRs start their journey across the Galaxy and eventually escape from the Galactic environment. In the standard picture used to describe Galactic propagation, CRs are produced by sources located in the thin Galactic disc and then diffuse in a larger volume where the presence of some magnetic turbulence is thought to scatter efficiently CRs. Beyond such {\it magnetic halo}, the turbulence is assumed to vanish such that particles can escape freely into the intergalactic space and the CR density reduces to $\approx 0$ \cite[(Berezinski \etal, 1990)]{Berezinski1990}.
The presence of a magnetic halo with a thickness $H$ much larger than the disc thickness, $h$, is required by the estimate of the residence time of CRs into the Galaxy which, assuming a diffusion coefficient $D(E)$ spatially constant across the whole halo, is given by the simple expression $\tau_{\rm esc}=H^2/D(E)$. Such quantity can be derived from the flux of radioactive nuclei present in the CR spectrum (like $^{10}$Be, $^{26}$Al and $^{36}$Cl, see \cite{Yanasak2001})  plus the ratio of secondary over primary (stable) nuclei (among which the one measured with better accuracy is B/C). Combining those information one gets the halo size $H\approx 3\div5$ kpc and the diffusion coefficient $D(R) \simeq 1\div5 \times 10^{28} (R/ \rm GV)^{-\delta}$ cm$^2$ s$^{-1}$, with $\delta \approx 1/3$, in good agreement with the Kolmogorov theory of turbulence \cite[(Aguilar \etal, 2015)]{Aguilar+2016}. Those values give $\tau_{\rm esc} \approx 150$ Myr for particle's rigidity around 1 GV
\footnote{Notice that the residence time estimated by \cite{Yanasak2001} using radioactive nuclei is only $\sim 15$ Myr. This result is obtained in the framework of the {\it leaky-box} model and is, to some extend, misleading in that it does not account for the spatial extention of the halo. It is known that when the decay time of unstable nuclei is smaller than the residence time, the leaky-box model gives incorrect results \cite[Berezinski \etal, 1990]{Berezinski1990}.}.

An independent proof to the existence of a magnetic halo comes from observations in the radio band which points towards the presence of diffuse synchrotron emission, revealing the presence of electrons and magnetic field above and below the galactic plane \cite[(Beuermann \etal,1985)]{Beuermann+1985}. The determination of the halo size in this case is not an easy task, partially due to the yet unknown structure of Galactic magnetic field. Nevertheless, comparison with numerical models for the CR electron distribution favors a halo size of $\sim 10$ kpc \cite[(Orlando \& Strong, 2013)]{OrlandoStrong:2013} while a size $\lesssim 2$ kpc seems to be strongly disfavored (\cite[Di Bernardo \etal, 2015]{diBernardo+2015}). It is worth mentioning that radio halos with a similar extension have been observed in others spiral galaxies observed edge-on(e.g. NGC 4631, NGC 891). 

Nevertheless, the halo model, at least in its basic version, does not provide a coherent physical picture. The value of $H$ required to explain the data is not well understood: it should be consistently determined by the spatial extension of the magnetic turbulence produced inside the Galactic disc (mainly by SN explosions), rather than used as a fit parameter.  Moreover it is difficult to imagine a sharp boundary where the particle transport abruptly changes from diffusion to free streaming regime and it is also unclear if such a boundary should be the same at all energies, as naively assumed. Finally,  the assumption that the diffusion coefficient should be equal everywhere in the halo is also a strong simplification.
The above criticisms are strengthened by the recent discovery of several anomalies both in the CR spectrum measured at Earth and in the diffuse Galactic $\gamma$-ray emission which both suggest a more complex situation with respect to the toy model depicted above. A detailed discussion of all these aspects can be found in \cite{AmatoBlasi:2017} while here we restrict our attention on two main results.

\begin{figure}
\begin{center}
 {\includegraphics[width=6.6cm]{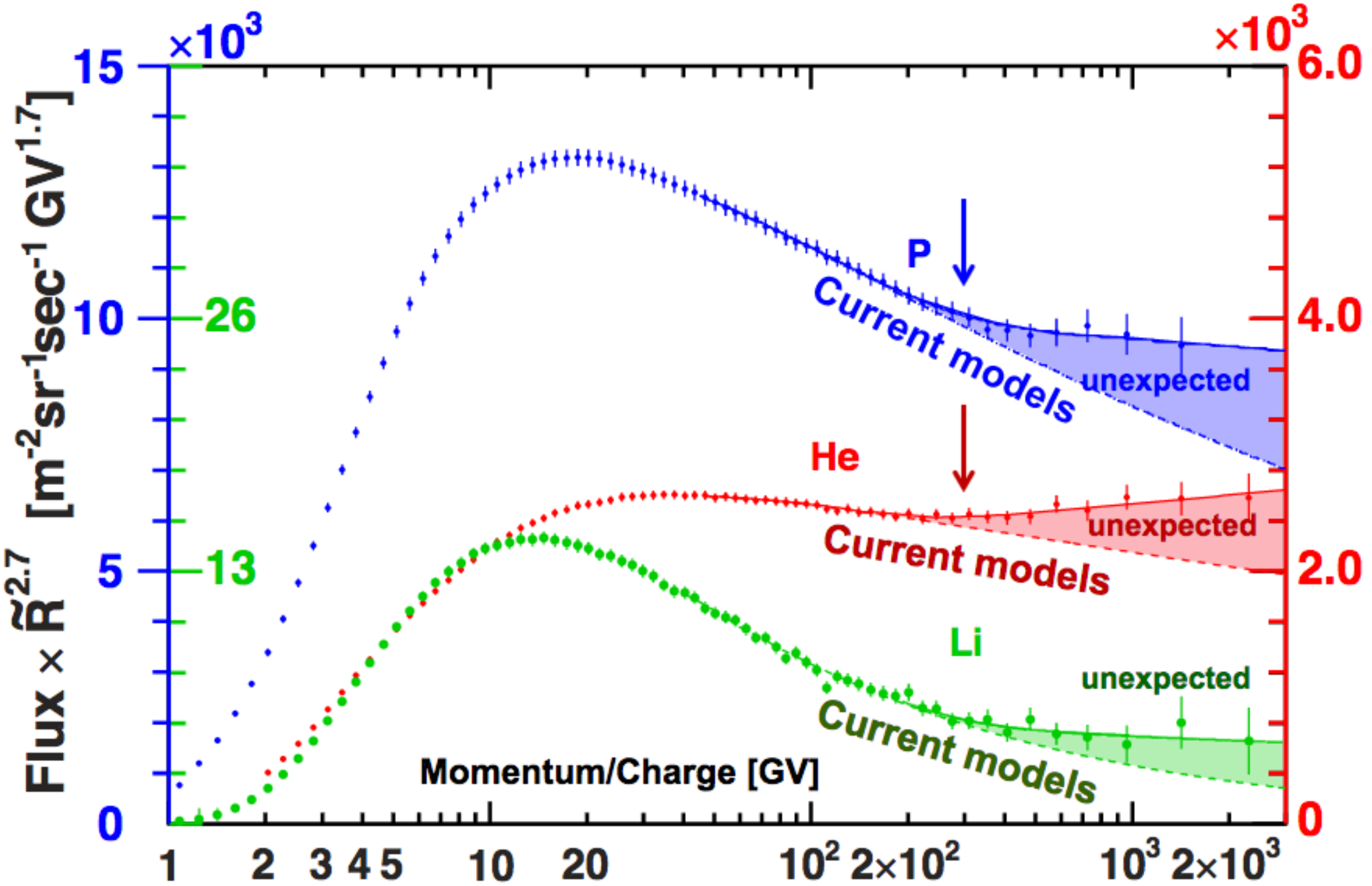} 
  \includegraphics[width=6.6cm]{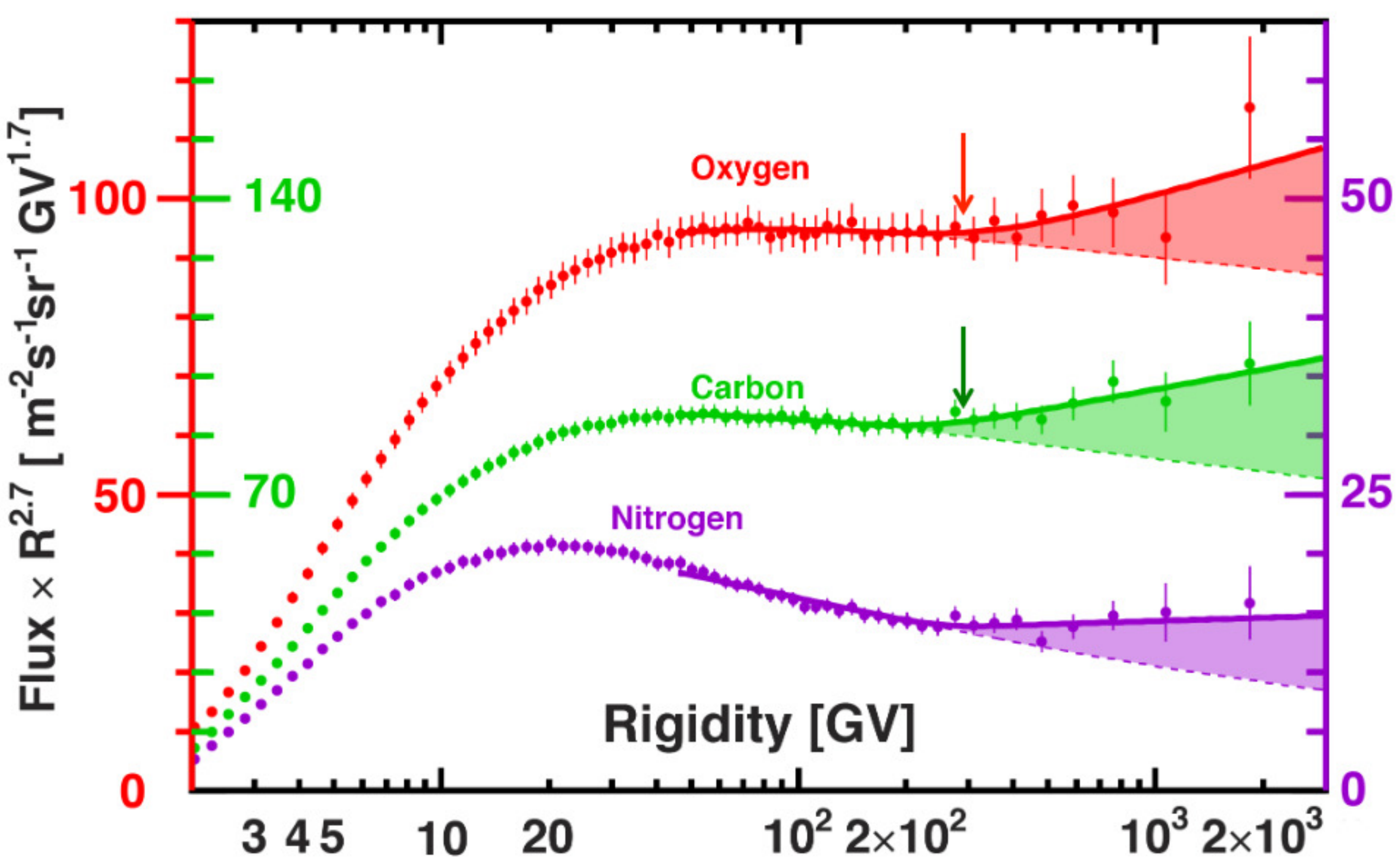} }
 \caption{Spectra of proton, Helium and Lithium (left plot) and Oxygen, Carbon, Nitrogen (right plot) as a function of rigidity. The shaded regions show the discrepancy above $\sim 300$ GV of the best fit models to data with the basic prediction of the standard model for CR propagation. Figures from \cite{Yan2017}.}
   \label{fig:CR_spectra}
\end{center}
\end{figure}

We start considering the anomalies in the local CR spectrum. Recent measurements carried out by PAMELA and AMS-02 have revealed that the spectra of protons and Helium are characterized by a spectral hardening above $\sim 300$ GV \cite[(Adriani \etal 2011; Aguilar \etal, 2015a,2015b)]{Adriani+2011, Aguilar+2015a, Aguilar+2015b}. Based on preliminary results by AMS-02, shown by  \cite{Yan2017}, a similar break might be present also in the spectra of heavier nuclei as shown in Fig.~\ref{fig:CR_spectra}. In standard approaches to CR transport, these findings can be accommodated imposing breaks in either the diffusion coefficient or the injection spectrum. We already discussed in the previous section a possible way to produce a break in the injected spectrum resulting from the superposition of two components: one due to particles escaping during the acceleration process and the other due to particles released after the SNR dissolution. Nevertheless, it is not clear if such a model remains viable once the integration over many SNRs is taken into account. 
Much attention has been payed to a possible modification of the transport process in the Galaxy. 
\cite{Dogiel+2002}, and more recently \cite{Tomassetti2012}, proposed a scenario where the diffusion coefficient changes with the distance from the Galactic plane. In particular, \cite{Tomassetti2012} discussed a two-zone halo model, where the dependence of the diffusion coefficient on the particle's energy is harder in the outer halo and softer to the inner one. Such a model can well explain the observed break in the CR spectrum, but remains only a phenomenological description, lacking of a more deep physical explanation of the halo's structure.
A more physically motivated scenario has been put forward by \cite{Blasi+2012, Aloisio&Blasi2013} and \cite{Aloisio+2015}, based on the idea that the magnetic turbulence could have two distinct components: one self-generated by the same CRs through the streaming instability and one resulting from the Kolmogorov cascade of turbulence injected at larger scale (presumably by SN explosions). In this scenario the diffusion is dominated at low energies by self-generated waves, while at larger energies, where the CR energy density becomes too low to generate a significant amount of waves, the Kolmogorov turbulence dominates and the spectral break in the CR spectrum should mark the energy where this transition occurs.

\begin{figure}
\begin{center}
 {\includegraphics[width=6.6cm]{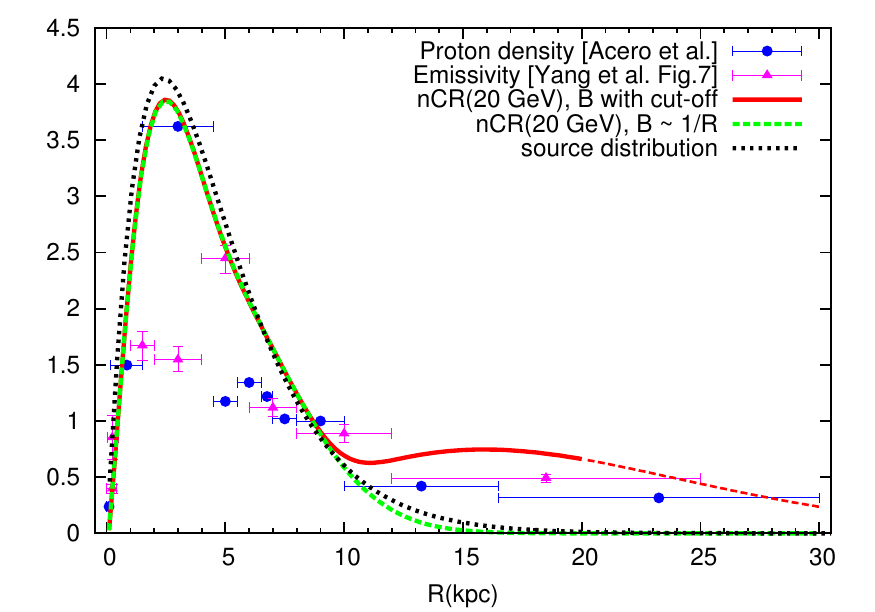} 
 \includegraphics[width=6.6cm]{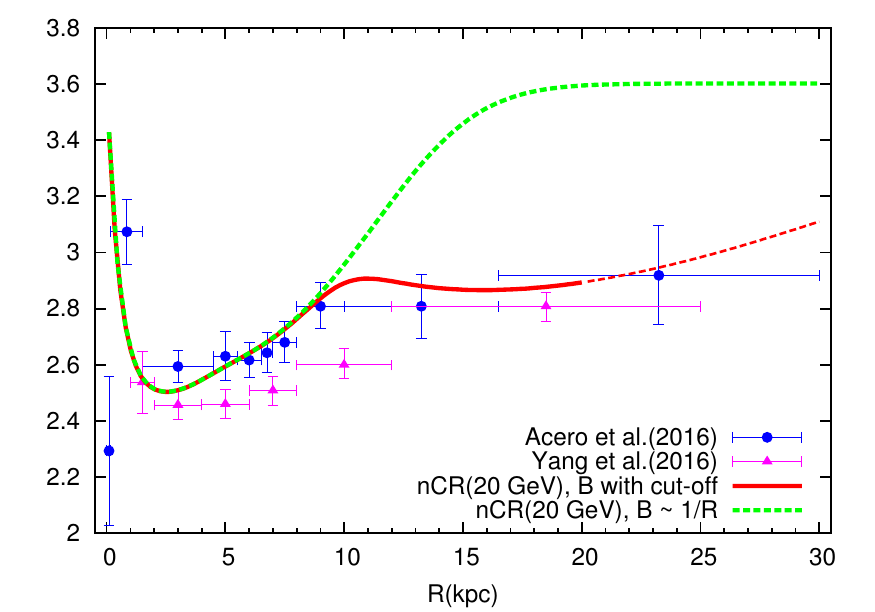}}
 \caption{CR density (left) and spectral slope (right) inferred from Fermi-LAT data as a function of galactocentric distance. Data correspond to two different analysis, \cite[(Acero \etal,  2016)]{Acero+2016} and \cite[(Yang \etal, 2017)]{Yang+2017}, as labelled. Solid-red and dashed-green curves show the CR density at $E > 20$ GeV (left) and the spectral slope (right) predicted by \cite{Recchia+2016a} using two different assumption for the Galactic magnetic field strength. Finally, the dotted-black line on the left shows the distribution of CR sources \cite[(Green, 2015)]{Green2015}.}
   \label{fig:CRgradient}
\end{center}
\end{figure}

One should keep in mind, however, that direct measurements can only probe the local CRs spectrum, hence the question whether such a spectrum reflects some peculiarities of the local environment, rather than being representative of the average Galactic CR spectrum, still remains open.
Relevant information on the CR distribution in the rest of the Galaxy can be obtained looking at the diffuse $\gamma$-ray emission which is thought to originate mainly from $\pi^0$ decay produced by collisions of CRs with the interstellar gas. Recent results presented by the Fermi-LAT collaboration \cite[(Acero \etal, 2017)]{Acero2017} and independently by \cite{Yang+2017}, show a substantial variation of the CR spectrum as a function of the distance from the Galactic Center. The spatial distribution of the CR density in the outer Galaxy appears to be weakly dependent upon the galactocentric distance, as found in previous studies as well, while the density in the central region of the Galaxy was found to exceed the value measured in the outer Galaxy. At the same time, Fermi-LAT data suggest a gradual spectral softening while moving outward from the center of the Galaxy to its outskirts, with a slope ranging from 2.6, at a distance of $\sim 3$ kpc, to 2.9 in the external regions. These findings represent a challenge for standard calculations of CR propagation based on assuming a uniform diffusion coefficient within the Galactic volume and may point towards a spatial-dependence of the CR transport properties, as already discussed by \cite{Evoli+2012} and \cite{Gaggero+2015}.
Indeed, this behavior could reflect, once again, the presence of self-generated turbulence which should dominate the particle scattering at low energies, the ones probed by Fermi-LAT. \cite{Recchia+2016a}, indeed, shows that the combination of self-generated turbulence plus the advection due to the wave motion at the Alfv\'en speed can account for both the spatial CR density and the spectral slope (see Figure~\ref{fig:CRgradient}). This result depends, to some extend, on the assumption of the Galactic magnetic field structure, which determines the effectiveness of the streaming instability. 

\underline{\it{Galactic winds}}.
We conclude this section discussing an interesting development of CR transport models that may have a large impact on the overall Galaxy evolution. Many galaxies produces a large scale wind that may affect star formation, through the regulation of the amount of gas available \cite[(Crain \etal, 2007; Stinson \etal, 2013)]{Crain+2007, Stinson+2013)} and inject hot gas in the galactic halo. Signatures for the presence of such hot gas have been presented also for the Milky Way through the detection of absorption lines and continuum emission in the X-ray band \cite[(Miller \& Bregman, 2015)]{MillerBregman2015}. 
Galactic winds are thought to be  powered by either SN explosions or starburst radiation \cite[(see e.g. Chevalier \& Clegg, 1985)]{ChevalierClegg1985}, but in our Galaxy the power injected by SNe and stars is too small to drive a large scale wind. 
On the other hand, the escape of CRs from the Galaxy inevitably leads to a gradient in the CR pressure that acts as a force on the background plasma, in the direction opposite to the gravitational pull \cite[(Ipavich \etal, 1975; Breitschwerdt \etal, 1991; Everett \etal, 2009)]{Ipavich1975, Breitschwerdt+1991, Everett+2009}. This force, added to the thermal pressure gradient, may contribute to lunch a galactic wind if the gas can be accelerated to supersonic speeds, otherwise the material can be lifted up and fall down in what is known as Galactic fountains. 
The existence of a Galactic wind would significantly change the transport properties of CRs. The main change is due to the presence of an advection speed, $u_{\rm adv}(z)$, which increases moving from the Galactic plane towards larger distances. As a consequence, above some distance, $s_*(E)$, given by the condition
\begin{equation}
  \frac{s_*^2}{D(E)} \approx \frac{s_*}{u_{\rm adv}(s_*)} \,,
\end{equation} 
the advection will dominate over diffusion and particles will be unable to diffuse back to the Galaxy. In other words, this distance $s^*$ represent an effective energy-dependent boundary which replaces the role of the halo size $H$. Such a model is particularly attractive because $s_*$ is not artificially imposed, like $H$, but it is rather self-consistently given by the wind solution \cite[(Ptuskin \etal, 1997)]{Ptuskin+1997}. 
Nevertheless, the wind model that includes only a diffusion coefficient due to self-generated turbulence, predicts a CR spectrum remarkably different from the one observed at Earth, especially at high energies where it is too steep. A better agreement with observations can be obtained assuming a pre-existing turbulence only in the near disc region \cite[(Recchia \etal, 2016b)]{Recchia+2016b}. 
In conclusions wind models seem to have a great potential to provide a comprehensive picture of the CR propagation, but at the moment such models are still immature and a lot of work needs to be done to include self consistently several relevant piece of physics like plasma cooling and heating processes, magnetic field damping and cascade of large scale turbulence. The complexity of such task will probably require time dependent simulations.

%SECTION%
\section{Conclusions} \label{sec:conc}
We presented a brief overview of some recent advancement in the field of CR physics, discussing separately the three main pillars that constitute the SNR paradigm for the origin of CRs: acceleration, escape and propagation.
Among the three, the acceleration mechanism is the most advanced field, being under investigation since the '70s. The non linear version of DSA, able to account for the back reaction of accelerated particles onto the shock structure, the magnetic field amplification, as well as the back reaction of the magnetic field itself, has provided a robust framework to interpret the rich phenomenology observed in SNRs. Several predictions has been verified, like, for example, the thin X-ray filaments observed in almost all young SNRs, interpreted as the result of strong magnetic field amplification produced by CRs.
In spite of these successes the theory is still unable to provide a firm interpretation of the particle spectrum inferred from non-thermal emission of SNRs, especially in the $\gamma$-ray band. In fact, the theory predicts particle spectra proportional to a power low in energy $E^{-2}$, with possible slight deviation due to non linear effects. On the contrary, observations shows spectral slope ranging from -1.5 to -2.5 for young remnants, while in the case of  middle aged SNRs the spectrum can be even steeper. This rich phenomenology is probably due to environmental effects, which are difficult to model. Moreover, the possible implications of those effects on the overall Galactic CR spectrum are, at the moment, unclear.

The escaping process has never been studied in great details, due to some theoretical difficulties but also to the lack of data coming from the escaping particles. A possible way to study this mechanism is through the $\gamma$-ray emission from SNR-MC associations. Few such complex have been already detected in $\gamma$-rays, but the future Cherenkov Telescope Array is expected to find tens of those systems, paving the way to a systematic study.

Finally the propagation process has received a lot of attention in the last few years, thanks to an avalanche of new data coming from both direct measurements (AMS-02 and PAMELA) and to indirect informations inferred from the diffuse $\gamma$-ray emission detected by Fermi-LAT. Thanks to those data, the basic propagation model start to shows its limitations, forcing the community to build more refined models. In particular I think we are close to a better understand of the magnetic turbulence that determine the CR propagation properties and, connected to this, also on the structure of the magnetic halo that surrounds the Galaxy.

It is worth mentioning that others anomalies found in the CR spectrum and not discussed here, like the rising positron fraction and the anti-proton flux, are driving new ideas that are questioning the basic pillars of the CR propagation \cite[(se, e.g., Katz \etal, 2010; Blum \etal, 2013; and Lipari, 2016)]{Katz+2010, Blum+2013, Lipari2016}. Whether these ideas can provide viable alternatives to the current paradigm for the origin of CRs is a matter of debate, which shows, anyhow, that the CR physics is undergoing a new phase of growing, potentially rich of new discoveries.

%\bibliographystyle{plain}
%\bibliography{biblio}

\end{document}